\documentclass[aps,prx,twocolumn,superscriptaddress,10pt]{revtex4-2}
\usepackage[colorlinks=true, allcolors=blue]{hyperref}
\usepackage{silence}
\usepackage[utf8]{inputenc}
\usepackage{amsmath}
\usepackage{amssymb}
\usepackage[normalem]{ulem}
\usepackage{graphicx}
\usepackage{diagbox}
\usepackage{pdfpages}
\makeatletter
\AtBeginDocument{\let\LS@rot\@undefined}
\makeatother
\usepackage{tikz}
\usepackage{mwe}
\usetikzlibrary{decorations.pathreplacing}
\usetikzlibrary{calc,intersections}
\usetikzlibrary{quantikz}
\usepackage{color}
\usepackage{verbatim}
\usepackage{booktabs}
\usepackage{float} 
\usepackage{listings}
\usepackage{alltt}
\usepackage{microtype}
\usepackage{hyperref}
\usepackage{cleveref}
\usepackage{subfigure}
\usepackage{xcolor}
\usepackage{outlines}
\usepackage{comment}

\usepackage{amsthm}

\newcommand{\cG}{{\mathcal{G}}}

\newcommand{\cO}{{\mathcal{O}}}

\newcommand{\invisible}[1]{}

\usepackage{ulem}

\begin{document}
\title{Low-depth random Clifford circuits for quantum coding against Pauli noise using a tensor-network decoder}
\date{\today}

\author{Andrew S. Darmawan}
\affiliation{Yukawa Institute of Theoretical Physics (YITP), Kyoto University, Kitashirakawa Oiwakecho, Sakyo-ku, Kyoto 606-8502, Japan}
\email[]{andrew.darmawan@yukawa.kyoto-u.ac.jp}
\affiliation{JST, PRESTO, 4-1-8 Honcho, Kawaguchi, Saitama 332-0012, Japan}
\author{Yoshifumi Nakata}
\affiliation{Yukawa Institute of Theoretical Physics (YITP), Kyoto University, Kitashirakawa Oiwakecho, Sakyo-ku, Kyoto 606-8502, Japan}
\author{Shiro Tamiya}
\affiliation{Department of Applied Physics, Graduate School of Engineering, The University of Tokyo, 7-3-1 Hongo, Bynkyo-ku, Tokyo 113-8656, Japan}
\author{Hayata Yamasaki}
\affiliation{Department of Physics, Graduate School of Science, The Univerisity of Tokyo, 7-3-1 Hongo, Bunkyo-ku, Tokyo 113-0033, Japan}
\affiliation{JST, PRESTO, 4-1-8 Honcho, Kawaguchi, Saitama 332-0012, Japan}

\begin{abstract}
Recent work [M.\ J.\ Gullans et al., Physical Review X, 11(3):031066 (2021)] has shown that quantum error correcting codes defined by random Clifford encoding circuits can achieve a non-zero encoding rate in correcting errors even if the random circuits on $n$ qubits, embedded in one spatial dimension (1D), have a logarithmic depth $d=\mathcal{O}(\log{n})$.
However, this was demonstrated only for a simple erasure noise model. In this work, we discover that this desired property indeed holds for the conventional Pauli noise model.  Specifically, we numerically demonstrate that the hashing bound, i.e., a rate known to be achieved with $d=\mathcal{O}(n)$-depth random encoding circuits, can be attained even when the circuit depth is restricted to $d=\mathcal{O}(\log n)$ in 1D for depolarizing noise of various strengths. This analysis is made possible with our development of a tensor-network maximum-likelihood decoding algorithm that works efficiently for $\log$-depth encoding circuits in 1D\@.
\end{abstract}

\maketitle
Protecting quantum information from noise and decoherence is a requirement for scalable quantum computation, which, in theory, can be done using quantum error-correcting codes.
Yet despite rapid developments in experimental realizations of quantum error-correcting codes~\cite{egan2021,Abobeih2022,Postler2022,acharya_suppressing_2022,PhysRevX.11.041058,Krinner2022,PhysRevLett.129.030501,10.1093/nsr/nwab011}, there are many challenges that must be overcome before they can be used in practical applications.

One major challenge is to deal with the daunting overhead required for various error correction schemes. For instance, experimental realization of the surface code is being pursued by several groups due to its high threshold and 2D layout \cite{dennis_topological_2002, fowler_surface_2012}. However, its major drawback appears to be a low encoding rate, defined as the $r:=k/n$ where $k$ and $n$ are the numbers of encoded and physical qubits of a quantum error correcting code, respectively. For practical computations, thousands of physical qubits may be required for each encoded qubit~\cite{litinski_magic_2019}. 

As a result, a significant effort has been made to find error correction schemes with a higher encoding rate and lower overhead.
A variety of schemes have been proposed based on quantum low-density parity-check (LDPC) codes~\cite{gottesman2014faulttolerant,PhysRevA.87.020304,8555154,panteleev_asymptotically_2022-1, leverrier_quantum_2022, gu_efficient_2022,leverrier_efficient_2022,https://doi.org/10.48550/arxiv.2206.07750,leverrier_parallel_2022} and concatenated quantum codes~\cite{https://doi.org/10.48550/arxiv.2207.08826}.
However, the existing high-rate quantum LDPC codes are hard to implement in many architectures since these codes require long-range two-qubit gates.
The high-rate concatenated code can suppress errors even if we can only use nearest-neighbor noisy two-qubit gates in 2D layout~\cite{https://doi.org/10.48550/arxiv.2207.08826,doi:10.1080/09500340008244046}, but may still need a higher circuit depth than is available under current technologies to attain sufficient error suppression.

Alternative approaches have been proposed, based on random encoding.
Random stabilizer codes are known to achieve a non-zero rate with vanishing error probability in the limit of large $n$ for certain types of noise~\cite{gottesman_stabilizer_1997, wilde_quantum_2013, NWY2021}. 
The rate achievable by random stabilizer codes against independently and ideally distributed (IID) Pauli noise $\mathcal{N}(\rho) = (1-p_I) \rho + p_X X \rho X + p_Y Y \rho Y + p_Z Z \rho Z$ is known as the hashing bound $r=1-H(p_I, p_X, p_Y, p_Z)$, where $X, Y, Z$ are Pauli matrices, $p_I, p_X, p_Y, p_Z$ represent Pauli error probabilities, and $H$ is the Shannon entropy~\cite{wilde_quantum_2013}.  While the hashing bound is not always optimal, it is relatively high compared to known upper bounds on the optimal rate~\cite{8046086,8119865}. It was shown by Brown and Fawzi~\cite{brown_short_2013, BF2013} that asymptotically the same performance can be achieved by random  Clifford encoding circuits even when the circuit depth is only $\mathcal{O}(\log^3 n)$. 

From a practical perspective, the above random codes have some shortcomings. In particular, an efficient decoding procedure for them is not known and they require all-to-all connectivity (which is not available in many physical architectures). A recent result of Gullans et al.~\cite{gullans_quantum_2021} has shown that random encoding by Clifford circuits with logarithmic depth and 1D connectivity or sub-logarithmic depth in higher dimensions can achieve a non-zero rate against erasure noise.
Similar performance was observed against erasure noise in an alternative construction of codes, based on constraint satisfaction algorithms~\cite{tremblay_finite-rate_2022}.  While restricting to the erasure noise model greatly simplifies the decoding of such codes, it involves the strong assumption that the locations of all physical errors are known,
 which is not the case in current quantum computing architectures. 
Hence, to be practically relevant, it is necessary to investigate the performance of such codes against more realistic noise models.

\begin{figure}
    \centering
    \includegraphics[width=0.4\textwidth]{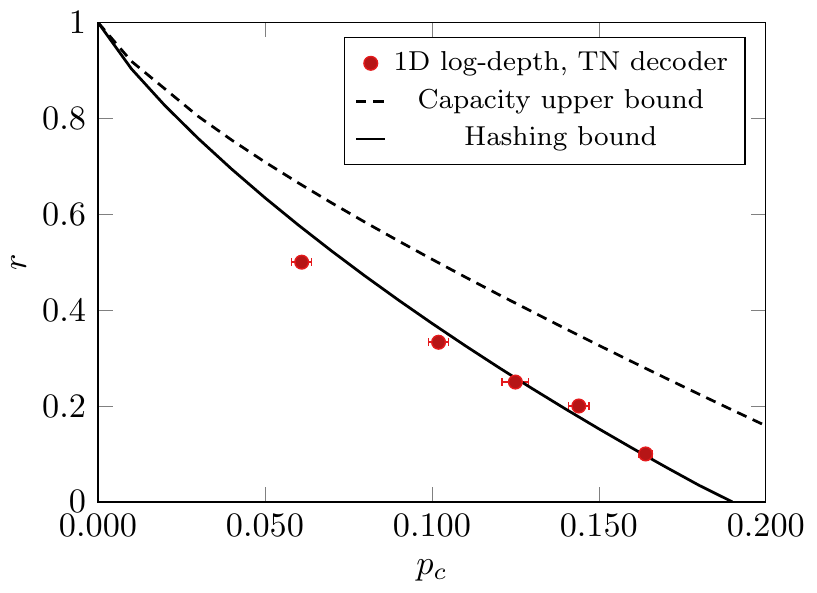}
    \caption{Numerical results for encoding rate $r$ vs threshold depolarizing probabilities $p_c$ of 1D log-depth circuits using a tensor network decoder (red markers) compared with various analytical bounds. The solid line is the hashing bound and the dashed line is an upper bound on the capacity, derived in Ref.~\cite{8119865}.}
    \label{fig:variousbounds}
\end{figure}

In this paper, we consider 1D random Clifford encoding circuits and demonstrate that, when the circuits have a logarithmic depth $d = \cO(\log n)$, the generated codes can be efficiently decoded and can  achieve a rate close to the hashing bound for depolarizing noise of various strengths. 
Our numerical thresholds are plotted alongside various analytical bounds in Fig. \ref{fig:variousbounds}.
We obtain our results by developing a tensor-network maximum-likelihood decoder for stochastic Pauli noise that has ${\rm poly} (n)$ running time when $d=\mathcal{O}(\log n)$. 
The combination of a high threshold against stochastic noise, non-zero rate, the practicality of low depth in 1D, and efficient decoding shows that such codes are promising candidates for quantum memories in future implementations of fault-tolerant quantum computers.

\par\textit{Low-depth random encoding circuits}\textemdash Here we briefly outline the definition of codes based on low-depth Clifford circuits. We start with a trivial quantum code with 1-qubit logical operators and checks. We partition the $n$ physical qubits into $k$ logical qubits and $n-k$ stabilizer qubits such that the logical qubits are evenly spaced among the physical qubits.
For each stabilizer qubit, indexed by $i\in \{1, \dots, n-k\}$, we randomly associate a non-trivial single-qubit Pauli check operator $g_i\in \{X, Y, Z\}$. The stabilizer of the code is the group $\mathcal{G}$ generated by the checks. The check operators trivially commute, and the code space is defined as the $+1$ eigenspace of all such check operators (or all elements of the stabilizer). This implies that the initial code space is a product state on the stabilizer qubits. The logical qubits, however, are not fixed by the checks, and for every logical qubit, indexed by $j\in\{1,\dots,k\}$, we  associate a pair of distinct anti-commuting single-qubit Pauli operators $l^x_j, l^z_j$, which we regard as the logical $X$ and logical $Z$ operators for that qubit.

Given a Clifford circuit $U$, we can produce a new stabilizer code by transforming the checks and logical operators as $g_i\mapsto U g_i U^\dagger$, $l^x_j\mapsto U l^x_j U^\dagger$ and $l^z_j\mapsto U l^z_j U^\dagger$. We assume $U$ to be noiseless. The circuit $U$ is an encoding circuit that maps unencoded logical qubits to encoded ones. The specific Clifford circuits we consider are low-depth circuits in 1D, where two-qubit iSWAP gates are applied in parallel between neighboring pairs of qubits in an alternating brickwork pattern (see Fig. 1(a) of Ref.~\cite{fisher_random_2022}). 
After each round of two-qubit gates, a uniformly random single-qubit Clifford gate is applied to every physical qubit. The depth $d$ of the circuit is taken to be the number of two-qubit gate layers. These locality constraints imply the weight of each check is at most $2d$.

For the TN decoder, which we define below, it is useful to consider open boundary conditions. To minimize the boundary effect, for a code with a given $n$ and $r$, we add an additional $4d-1/r+1$ stabilizer qubits to make sure all logical qubits are at least $2d$ physical qubits away from the boundary before applying the encoding circuit. This  changes the rate of the code, however, given that we restrict to $d=\cO(\log n)$, this does not affect the asymptotic rate as $n\rightarrow \infty$. In a slight abuse of notation, we use $r$ to refer to the rate of the code before the boundary qubits are added and we define $n_{\rm phys}:=n+4d-1/r+1$ to be the total number of physical qubits (including boundary qubits) of the code.

\par\textit{Tensor-network decoding}\textemdash
To assess the performance of these codes, we consider the following scenario. First, the logical information is encoded into an error-correcting code using the encoding circuit $U$ as described above. Next, every physical qubit suffers noise, which we assume is depolarizing noise. Finally, all of the checks are measured, and decoding is performed. We assume the measurements are noiseless. 
Decoding is a classical computation that takes the check measurement outcomes, called the syndrome, as input and outputs a correction to restore the encoded data. For stochastic Pauli noise, this decoding task (a classical computational problem) is hard in general~\cite{iyer_hardness_2015}; however, we show that efficient near-optimal decoding is possible for 1D random Clifford codes of depth $d = \mathcal{O}(\log n)$. 
Note that, while this scenario is not fully realistic due to the assumptions of noiseless encoding circuits and measurements, these assumptions are conventionally used to quantify code performance in a code capacity setting~\cite{https://doi.org/10.48550/arxiv.1108.5738}.
\begin{figure}
    \centering
    \includegraphics[width=0.46\textwidth]{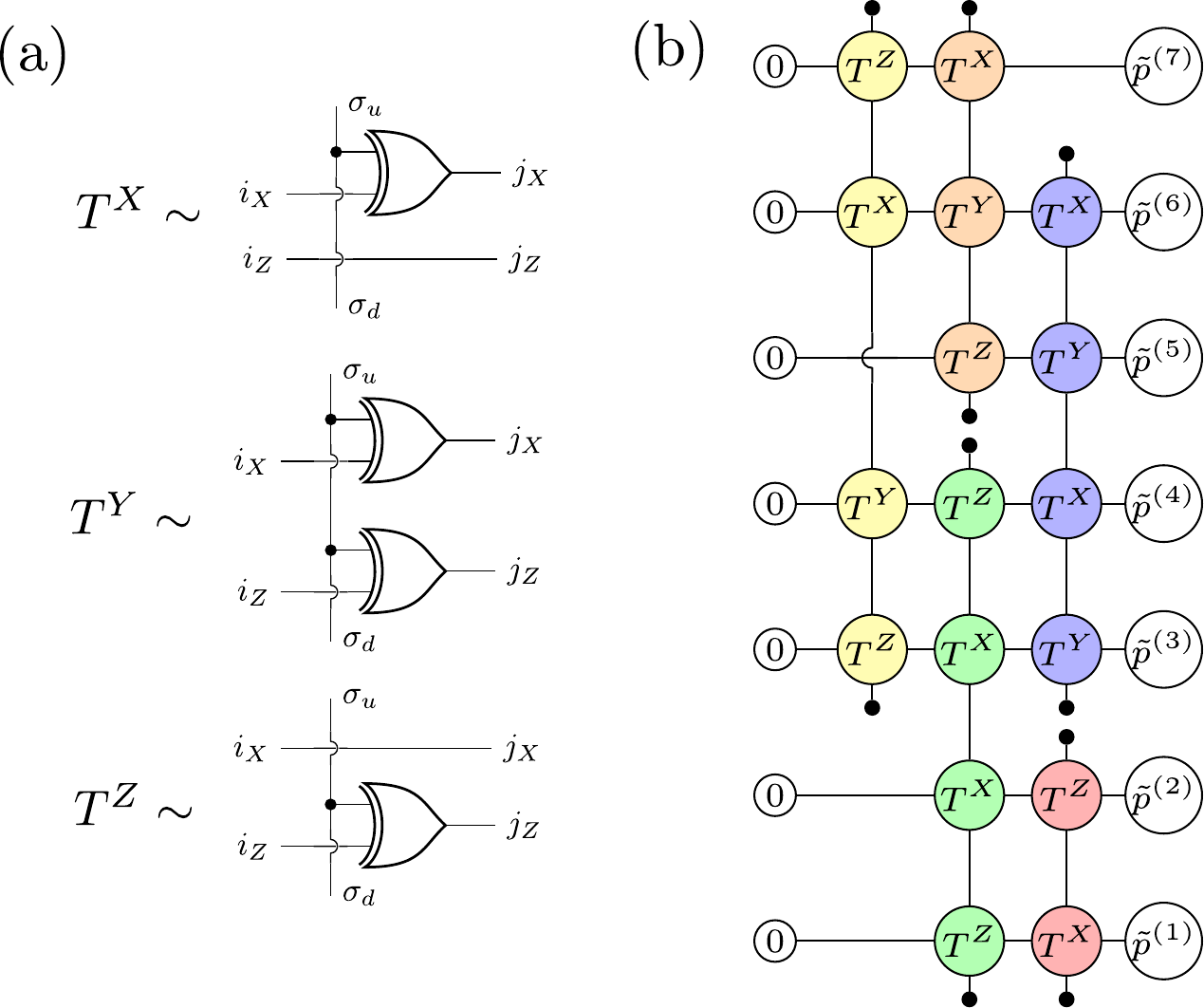}
    \caption{(a) Tensors defined in terms of logic circuits. The entries of these tensors are 1 for any assignment of bits to the tensor indices that are valid with respect to the circuit, and 0 otherwise. In particular, it is 1 only if $\sigma_u = \sigma_d$ for each tensor. The logic gate appearing in these circuits is the XOR gate. In computing $p(f_Le)$, where $e \in \cG$, the indices $\sigma_u$ and $\sigma_d$ specify if a stabilizer generator is contained in $e$ or not, and $(i_X, i_Z)$ and $(j_X, j_Z)$ are for keeping track of changes of Pauli operators if the generator is applied.
    (b) The coset probabilities for maximum-likelihood decoding for any stabilizer code can be expressed as the contraction of a two-dimensional TN\@. Each horizontal row corresponds to a physical qubit of the code, so the number of rows is always $n_{\rm phys}$. The network is constructed from generators of $\cG$ (checks). In this illustration, nodes corresponding to different generators are distinguished by their color. As illustrated, the generators are sorted into columns and are arranged such that no two generators overlap in a given column. Non-trivial Pauli operators in a generator are replaced with the corresponding tensors in (a), and all tensors in a generator are connected by a vertical wire. 
    This implies that all $\sigma_u$ and $\sigma_d$ in the tensors corresponding to a given check must coincide to contribute a non-zero term to the contraction. 
    The tensor $\tilde{p}^{(j)}$ is a vector of the four Pauli error probabilities on site $j$ which are permuted according to the action of $f_L$ on site $j$. The 0-tensors on the left fix the left indices to 0, and the small black tensors have entries all equal to 1.
    By horizontally contracting the tensors in a given row with $\sigma$ indices fixed, either $p_I, p_X, p_Y$ or $p_Z$ on the $j$th qubit is obtained depending on what Pauli operators act on the qubit in $f_Le$. Contracting the network corresponds summation over all $\sigma$ indexes, and thereby all $e\in \cG$, which evaluates to $p(f_L \cG)$ as in Eq. \eqref{e:coset_prob}.}
    \label{fig:tensornet}
\end{figure}

Here we briefly outline how the decoding problem can be cast as a tensor network (TN) contraction. Full details are provided in Appendix \ref{s:tn_decoder}. Say that all the checks are measured, and the syndrome $s$ is obtained, and let $f$ be a Pauli operator that anti-commutes only with the flipped checks (such a Pauli operator can always be computed efficiently by row-reduction of the check matrix given $s$).
Due to the fact that two Pauli errors have an identical effect on the code if and only if they differ by multiplication by an element of the stabilizer,
the probability that $f$ will correct the physical error $e_p$ is the probability that $e_p\in f\mathcal{G}$, which is given by 
\begin{equation}
p(f\mathcal{G})=\sum_{e\in \mathcal{G}} p(fe)\,, \label{e:coset_prob}
\end{equation}
where $p(fe)$ denotes the probability of Pauli error $fe$.
For independent Pauli noise, each $p(fe)$ is a product of Pauli error probabilities. If $p(f\mathcal{G})$ can be computed for any $f$, then optimal, maximum-likelihood decoding can be performed by computing this probability for all logically inequivalent corrections $f_L := fL$ consistent with the syndrome, where $L$ are logical operators, and choosing $f_L$ with the largest probability among the set $\{p(f_L \mathcal{G})\}_L$.
However, it is inefficient to compute the probabilities directly by evaluating the sum in Eq. \eqref{e:coset_prob} since the number of terms in the summation is $|\mathcal{G}| = 2^{n-k}$. In fact, we do not expect an efficient algorithm to exist for computing coset probabilities of codes in general, due to the \#P-hardness of the problem~\cite{iyer_hardness_2015}.

Fortunately, for codes with local checks, there are examples where coset probabilities can be computed efficiently using TN methods. We present two such methods for evaluating coset probabilities that are efficient and nearly optimal when restricted to 1D codes with $\log$-depth encoding circuits. One of them, which we describe in Appendix~\ref{s:basic_decoder}, is based on the TN description of coset probabilities for the surface code in Ref.~\cite{bravyi_efficient_2014}.

The results presented in this paper, however, have been computed using an alternative TN description of the coset probabilities, which we illustrate in Fig.~\ref{fig:tensornet}. The caption sketches the TN and briefly explains how it works. See Appendix \ref{s:new_tn} for the full details.
In the case of codes with 1D encoding circuits of depth $d$, the shape of the TN is $n_{\rm phys}\times \mathcal{O}(d)$. When we restrict to low depth $d=\mathcal{O}(\log n)$, such a TN can be contracted exactly in ${\rm poly}(n)$ time.
Note that this is in contrast to the problem of contracting an $n\times n$ square lattice TN, which is known to be \#P-complete~\cite{schuch_computational_2007}.

To contract the TN, we have used a simple exact contraction algorithm. However, its structure as a square lattice TN leaves potentially many alternative approximate or exact contraction schemes for speeding up the contraction~\cite{pan_simulation_2022}. We expect that the TN description based on logic circuits is likely most suited to codes where the check weight is relatively large since the size of individual tensors does not grow with the check weight.

While each coset probability can be computed exactly and efficiently with this method, an approximation is used due to the fact that the total number of inequivalent cosets to be maximized over is $4^k$. As described in more detail in Appendix~\ref{s:marginal}, we overcome this by decoding each logical qubit independently while marginalizing over other qubits. This requires only $4k$ coset probabilities to be computed and maximized over in total but could be suboptimal if logical failures are highly correlated (however, this does not seem to be the case, as shown Appendix \ref{s:short_range}).
\begin{figure}[t]
    \centering
    \includegraphics[width=0.4\textwidth]{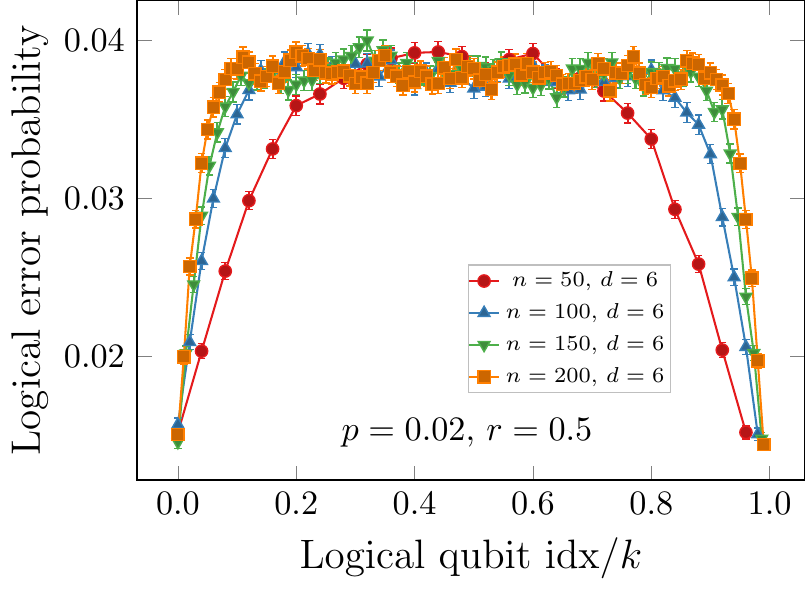}
    \caption{Error probability of a logical qubit vs. logical qubit index divided by $k$ (corresponding to its relative position on the chain) for $r=0.5$ with $d=6$ and $p=0.02$. The boundary effect can clearly be seen. The error rate stabilizes to a system-size independent value at a certain distance from the boundary. }
    \label{fig:basic_properties}
\end{figure}

\par\textit{Numerical results}\textemdash
We have performed simulations using the TN decoder described above to study the properties of codes defined by random Clifford encoding circuits in 1D\@. 
In each run of the simulation, we randomly generate a code using a Clifford circuit of depth $d=\mathcal{O}(\log n)$ as described above and sample a Pauli error $e_p$ according to the depolarizing noise model, which gives rise to a syndrome $s$. The decoder calculates a correction $f_L$, using $s$ as input.

We say that logical qubit $j$ fails when $f_Le_p$ anticommutes with at least one of the logical generators $Ul_j^xU^{\dagger}$ or $Ul_j^zU^{\dagger}$, which can be easily checked. 
At least $2\times 10^5$ runs of the simulation are performed for each data point. Taking the average over the stochastic Pauli noise as well as over the random Clifford encodings, we can estimate the probability of any given logical qubit failing. 

In Fig.~\ref{fig:basic_properties}, we show the spatial distribution of logical errors for various system sizes. Although there is a clear boundary effect, we observe that the logical failure probability for qubits sufficiently far from the boundary is uniform across logical qubits and independent of system size if $d$ is kept constant. We let $p_L'$ denote the failure probability of a logical qubit in the bulk region. 
In Appendix \ref{s:short_range}, we present numerical results that indicate, unlike the fully random codes, the spatial correlations of logical errors are short-ranged, as observed in Ref.~\cite{gullans_quantum_2021} for erasure noise.

\begin{figure}
    \centering
    \includegraphics[width=0.4\textwidth]{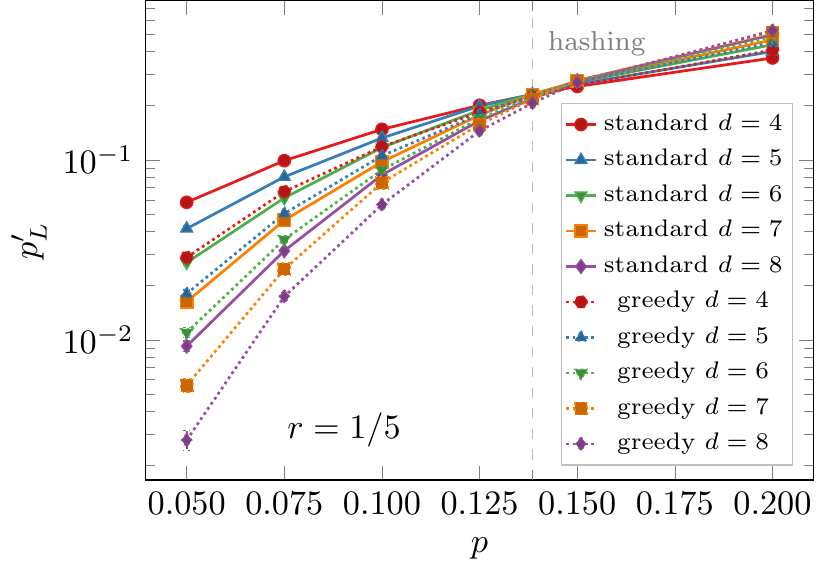}
    \caption{Bulk logical error probability $p_L'$ versus physical error probability $p$ for $r=1/5$ using a fixed system size of $n=50$, excluding $\mathcal{O}(d)$ added boundary qubits using both standard and greedy random encoding circuits.  A clear crossing point can be observed, which is very close to the threshold error probability implied by the hashing bound $p=0.139$ indicated by the dashed grey line. Computed thresholds and threshold plots for other rates are included in Appendix \ref{s:supplementary_numerics}.}
    \label{fig:thresholds}
\end{figure}

We have plotted $p_L'$ as a function of the physical error probability for a variety of depths $d$ and $r=1/5$ in Fig.~\ref{fig:thresholds}.  The crossing point for curves of different $d$ indicates a threshold in the code, below which $p_L'$ decays exponentially in $d$. Evidence of this exponential decay, plots at other rates, as well as a list of computed thresholds, are included in Appendix \ref{s:supplementary_numerics}. The numerically obtained thresholds are plotted alongside the hashing bound in Fig. \ref{fig:variousbounds}.

The performance of the code can be improved by slightly modifying the random Clifford circuits. As one instance, we propose a random `greedy' code by choosing single qubit gates in the random circuit to maximize the weight of checks and logical generators in each layer, rather than uniformly at random. See Appendix \ref{s:improvements_random} for the details. The results of the greedy construction are also provided in Fig.~\ref{fig:thresholds} alongside the standard construction.
 Despite differences in logical error rates, the threshold crossing points are very similar, the greedy construction appears to behave like the standard construction except with a greater effective depth.
 
 For both constructions, these plots show that the thresholds are very close to the hashing bound for $r\lesssim 1/3$. This is remarkable since these codes achieve the same threshold as a fully random code for these rates, despite being much more local and restricted. At higher rates, e.g.\ $r=1/2$, a threshold is harder to discern from the data; however, we conjecture that using larger values of $d$ than is currently accessible with our numerical method, would produce a threshold estimate close to the hashing bound, as with the lower rates. 

\begin{figure}
    \centering
    \includegraphics[width=0.4\textwidth]{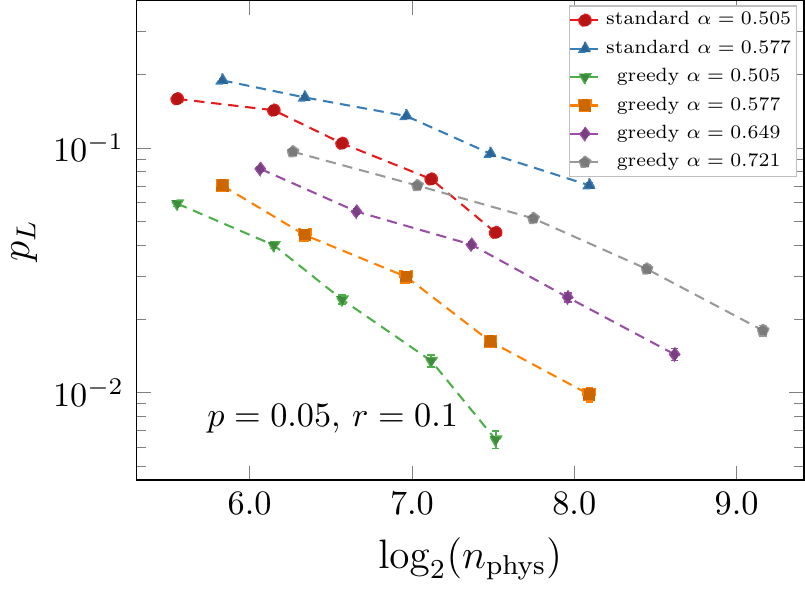}
    \caption{Probability of at least one logical qubit failing vs $n_{\rm phys}$ when $d=\alpha^{-1}\log_2{k}$, for various $\alpha$. The error rate decays to zero with $n_{\rm phys}$, and by varying $\alpha$ we can increase the number of encoded qubits at the cost of a slower rate of decay.}
    \label{fig:large_codes}
\end{figure}

Note that the observed exponential rate of decay in $d$ of $p_L'$ below the threshold for $r\lesssim 1/3$ implies that the probability of an error occurring on at least one logical qubit, which we refer to as $p_L$, also tends to zero as $n\rightarrow \infty$ provided that $\log_2 k = \alpha d$ for a sufficiently small constant $\alpha>0$. In other words, $d=\cO(\log n)$ for fixed $r$ can be sufficient for $p_L$ to tend to zero as $n\rightarrow \infty$. Note that $1-p_L$ is equal to the entanglement fidelity of the $k$-qubit logical channel (which is proportional to the average channel fidelity~\cite{nielsen_simple_2002}).
In Fig.~\ref{fig:large_codes}, we plot $p_L$ as a function of $n_{\rm phys}$ with fixed rate $r=0.1$ and $p=0.05$ and various $\alpha$. As can be seen, by varying $\alpha$, we observe a trade-off between the rate of decay of the total logical error rate and the total number of encoded qubits $k=rn$. For practical error correction, it would likely be useful to tune the value of $\alpha$ as well as the rate $r$, according to the target number and error probability of logical qubits. 

\par\textit{Discussion}\textemdash
In this work, we have studied quantum error correcting codes defined by 1D low-depth Clifford encoding circuits.
We have shown that for the family of these codes with logarithmic depth $d=\mathcal{O}(\log n)$, maximum-likelihood decoding of stochastic Pauli noise can be performed in ${\rm poly}(n)$ time using TN methods.
We have also numerically shown that, for depolarising noise over a large range of noise strengths, the codes can achieve a rate close to the hashing bound if $d = \cO(\log n)$. 
Thus, 1D Clifford encoding circuits with depth $\cO(\log n)$ can generate quantum error correcting codes that have the same rate as random stabilizer codes and can be efficiently decoded.
The high performance, non-zero rate as well as locality
in 1D suggest that such codes could serve as practical quantum memories in future implementations of quantum computers. 

The results suggest a number of potential directions for future research. Firstly, it would be interesting to see whether codes defined by low-depth Clifford encoding circuits in two or higher dimensions have advantages over the 1D codes considered here. 
This was shown to be the case for the erasure noise model when the code is modified with a process called expurgation~\cite{gullans_quantum_2021}. Whether the same holds against Pauli noise remains unknown.

Another direction towards practical implementation is to consider the realistic case where syndrome extraction is itself prone to error. In this case, there is the added complexity that higher-weight stabilizer measurements are likely to be less reliable than low-weight ones. Furthermore, fault-tolerant methods for state preparation and logical gates must be developed. On potential route is via Knill's fault-tolerant error correction gadgets that work for any stabilizer code even with a non-constant-weight stabilizer like ours~\cite{knill2005quantum,https://doi.org/10.48550/arxiv.quant-ph/0402171}.

The above directions for future research may require generalizations to the TN decoding methods described in this work. 
For this, one could consider leveraging more advanced TN contraction methods or adapting non-TN decoding methods that have been applied to other types of codes.

Finally, while we believe that these numerical results strongly suggest that the rate of these codes is close to the hashing bound, they do not constitute mathematical proof. It would be interesting to see whether the proofs supporting, or contradicting these suggestions can be made by, for instance, strengthening bounds in Ref.~\cite{BF2013}. 

\par\textit{Acknowledgments}\textemdash
ASD was supported by JST, PRESTO Grant Number JPMJPR1917, Japan. YN is supported by MEXT-JSPS Grant-in-Aid for Transformative Research Areas (A) ``Extreme Universe”, Grant Numbers JP21H05182 and JP21H05183, and by JSPS KAKENHI Grant Number JP22K03464.
ST was supported by JST Moonshot R\&D Grant Number JPMJMS2061.
HY was supported by JST PRESTO Grant Number JPMJPR201A and MEXT Quantum Leap Flagship Program (MEXT QLEAP) JPMXS0118069605, JPMXS0120351339\@.
The numerical computation in this work was carried out at the Yukawa Institute Computer Facility.

\bibliographystyle{apsrev4-2}
\bibliography{physics.bib}

\appendix

\section{TN descriptions of coset probabilities}
\label{s:tn_decoder}
Here we provide full details of the TN decoder which we have used to study codes produced by low-depth random Clifford circuits in 1D. We show that the TN decoder runs in polynomial time for codes with 1D log-depth encoding circuits. 

Say an $n$-qubit Pauli error $e_p$ occurs on a state in the code space, and all of the checks are measured, yielding the syndrome outcomes $s=s_1, s_2, \dots, s_{n-k}$ where $s_i\in \{-1,1\}$ is the  outcome of the measuring check $g_i$.
Let $f$ be any product of Pauli operators that is consistent with that syndrome, in that it anti-commutes with the checks that returned a ${-}1$ outcome and commutes with the other checks. The physical error $e_p$ is one such error, but it is not known to the experimenter. One can find an operator consistent with any syndrome by performing row-reduction on the check matrix, the rows of which are binary vectors representing the check operators. 

The operator $f$ applied to the code will correct $e_p$ if $e_p \in f \mathcal{G}$. If not, then $fe_p=L$ for some non-trivial logical operator $L$, and $fLg$ for any $g\in \mathcal{G}$ will correct the error. Maximum-likelihood decoding finds the correction that is most likely to correct the error. To do this, we determine the $L$ for which $p(fL\mathcal{G})$ is maximized, i.e., the coset $fL\mathcal{G}$ that $e_p$ most likely belongs to. For any $L$, we henceforth combine the $f$ and $L$ into a single Pauli error $f_L$.

The probability of an error belonging to the coset $f_L\mathcal{G}$ is simply the sum of the probabilities of every error in that coset i.e. $p(f_L\mathcal{G}) = \sum_{e\in \mathcal{G}} p(f_Le)$. By the definition of the generating set, every element of $\mathcal{G}$ is of the form $e(\sigma)=\prod_{i=1}^{n-k} g_i^{\sigma_i}$, with $\sigma=\sigma_1, \sigma_2, \dots \sigma_{n-k}$,  $\sigma_i\in\{0,1\}$ for each $i$ representing a particular check configuration.

For independent Pauli noise, we can write the probability of an error $f_Le(\sigma)$ as a product of single-qubit Pauli error probabilities $p(f_Le(\sigma))=\prod_{j=1}^n A^{(j)}_\sigma(L)$, where $A^{(j)}_\sigma(L)\in\{p^{(j)}_I, p^{(j)}_X, p^{(j)}_Y, p^{(j)}_Z\}$ for each $\sigma$ and where $p^{(j)}_Q$ is the probability of Pauli error $Q$ occurring on qubit $j$. 
Note that, if the noise is identical for each qubit $j$,  $p_P^{(j)}$ ($P=I,X,Y,Z$) does not depend on $j$.
Since this discussion will mainly focus on the computation of a single coset (with fixed $L$), we will henceforth drop the dependence of $A^{(j)}_\sigma(L)$ on $L$ from our notation.

While $A^{(j)}_\sigma$ for a particular site $j$ depends on the check configuration $\sigma$, it is clear that it only depends on the bits $\sigma_i$ for which $g_i$ acts non-trivially on qubit $j$. Thus, each term $A_\sigma^{(j)}$ can be written as a tensor of rank $r_j$, where $r_j$ corresponds to the number of generators that act non-trivially on $j$. 
For a general stabilizer code, we replace $A_\sigma^{(j)}$ with $A_{\sigma(j)}^{(j)}$, where $\sigma(j)$ is the list of all check bits $\sigma_i$ for which $g_i$ acts non-trivially on $j$. 

To obtain the coset probability, we sum over all indices 
\begin{equation}
   p(f_L\mathcal{G})=\sum_{\sigma}\prod_{j=1}^n A^{(j)}_{\sigma(j)}\,,
   \label{e:summation}
\end{equation}
Note that the expression on the right-hand side has a similar form to a TN contraction. Each $A^{(j)}_{\sigma(j)}$ is a tensor of rank $r_j$, and the sum of the product structure is nearly identical to tensor contraction. One small difference is that each index $\sigma_i$ can appear in more than two tensors (they essentially correspond to hyperedges of the TN, compared to usual graph edges, that only connect two nodes/tensors). We can convert this into a typical TN in the following ways. 

\subsection{TN description of coset probabilities, based on Ref.~\cite{bravyi_efficient_2014}}
\label{s:basic_decoder}

One way to turn the summation in Eq.~\eqref{e:summation} into a TN, as done in Ref.~\cite{bravyi_efficient_2014}, involves adding a $\delta$ tensor for each check, with only two non-zero entries specified by $\delta_{\sigma_1, \sigma_2, \dots, \sigma_n}=1$ if $\sigma_1=\sigma_2=\dots=\sigma_n= 0 {\,\,\rm or\,\,} 1 $. The TN is defined by connecting indices of $A$ tensors to the appropriate check tensors $\delta$. 

\begin{figure}
    \includegraphics[width=0.45\textwidth]{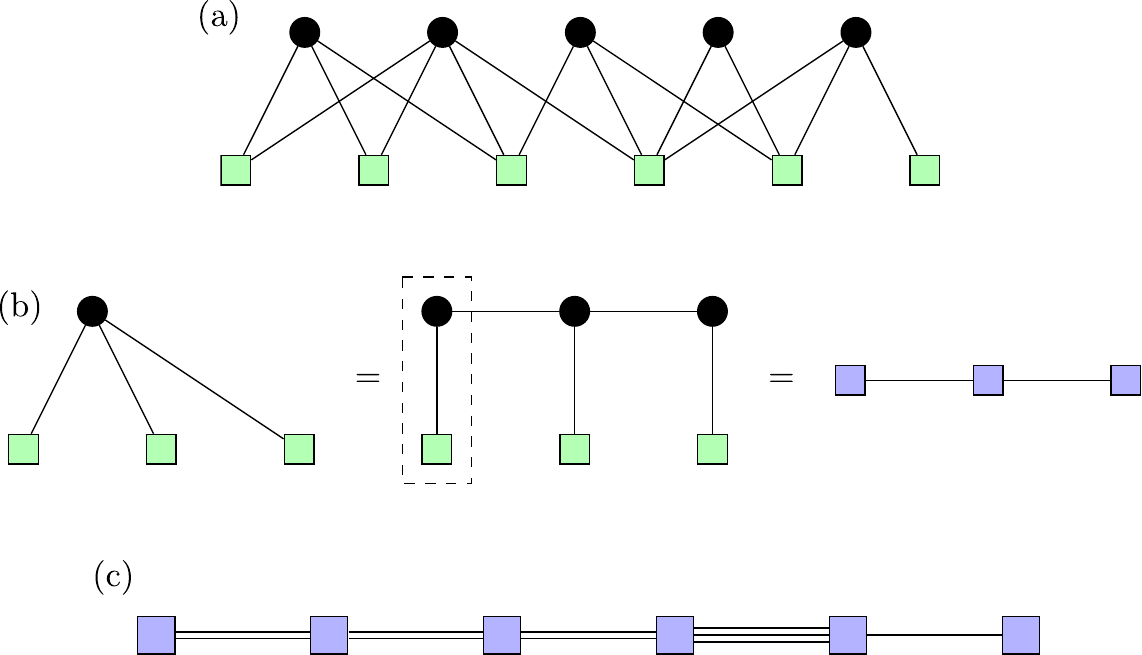}
    \caption{(a) Coset probabilities represented as a TN following the construction in Ref.~\cite{bravyi_efficient_2014}. Each black circle node corresponds to a check and each green square node corresponds to a qubit. An edge is drawn between a check node and a qubit node if and only if the check acts non-trivially on that qubit. The check nodes correspond to $\delta$ tensors and the green nodes correspond to $A_{\sigma(j)}$ tensors (defined in the text). (b) $\delta$ tensor can be split and combined with the qubit tensors, resulting in the blue square tensors. (c) This is done for every $\delta$ tensor in (a), resulting in a 1D TN. For 1D codes with depth $\mathcal{O}(d)$ encoding circuits, the maximum number of edges connecting a pair of neighboring tensors is $\mathcal{O}(d)$, resulting in a maximum matrix size of $2^{\mathcal{O}(d)}\times 2^{\mathcal{O}(d)}$.}
    \label{fig:bsv_tn}
\end{figure}

The TN then has the same form as the Tanner graph of the code, as illustrated in Fig.~\ref{fig:bsv_tn}. This is a bipartite graph with two types of nodes, which we refer to as qubit nodes and check nodes. Each qubit node corresponds to a physical qubit and each check node corresponds to a code check. An edge is added between a qubit node and a check node if and only if the check acts non-trivially on the corresponding qubit. This graph is mapped to a TN by placing a $\delta$ tensor at every check node and an $A$ tensor at every qubit node.

This TN description has proven useful for the surface code and other planar codes~\cite{chubb_general_2021} since it can be contracted efficiently using established TN methods. However, for families of codes whose check weight grows with $n$, like the random codes studied in this paper, one encounters the problem that the size of the tensors grows exponentially with the weight of the checks. Furthermore, the resulting TN is not planar and so the methods of Ref.~\cite{chubb_general_2021} cannot be applied directly. 

Nevertheless, for 1D Clifford encoding circuits, the TN can be converted into a 1D TN as shown in Fig.~\ref{fig:bsv_tn}. The 1D TN consists of a product of (sparse) matrices of size $2^{\mathcal{O}(d)}\times 2^{\mathcal{O}(d)}$. When $d=\mathcal{O}(\log n)$, this TN can be contracted exactly in polynomial time in $n$. 

We have implemented this decoder and confirmed that it produces the correct coset probabilities. We did not use this TN description to obtain the results in this paper, since our implementation of it was slower than the method described in the following section. We note, however, that this description could prove useful if optimized e.g.\ by exploiting the sparse structure of the matrices. For the remainder of this section, we focus on the following alternative TN of coset probabilities.

\subsection{Alternative TN description of coset probabilities}
\label{s:new_tn}
Here we describe an alternative way to efficiently represent the coset probabilities $p(f_L \mathcal{G})$ for any stabilizer code as a two-dimensional network of small tensors. We have used this method to produce the results presented in this paper. The network is illustrated in Fig.~\ref{fig:tensornet} in the main text. 

Most of the tensors in the network are from the set $\{T^X, T^Y, T^Z\}$, which are defined in Fig.~\ref{fig:tensornet}(a) in terms of classical logical circuits. The tensors have an entry $1$ for each index assignment corresponding to a valid execution of the circuit, and the remaining entries are zero. 
The tensors are precisely defined as follows: for $P=X, Y, Z$, $T^P_{i_X, i_Z, j_X, j_Z, \sigma_u, \sigma_d} = 0$ if $\sigma_u \neq \sigma_d$,
\begin{align}
    &T^P_{i_X, i_Z, j_X, j_Z, 0, 0}
    = 
    \begin{cases}
    1 & \text{if $(i_X, i_Z) = (j_X, j_Z)$},\\
    0 & \text{otherwise},
    \end{cases}
\end{align}
and
\begin{align}
    &T^X_{i_X, i_Z, j_X, j_Z, 1, 1}
    = 
    \begin{cases}
    1 & \text{if $(i_X, i_Z) = (j_X \oplus 1, j_Z)$},\\
    0 & \text{otherwise},
    \end{cases}\\
    &T^Y_{i_X, i_Z, j_X, j_Z, 1, 1}
    = 
    \begin{cases}
    1 & \text{if $(i_X, i_Z) = (j_X \oplus 1, j_Z \oplus 1)$},\\
    0 & \text{otherwise},
    \end{cases}\\
    &T^Z_{i_X, i_Z, j_X, j_Z, 1, 1}
    = 
    \begin{cases}
    1 & \text{if $(i_X, i_Z) = (j_X, j_Z \oplus 1)$},\\
    0 & \text{otherwise}.
    \end{cases}
\end{align}

As we will see below, in the computation of the probability $p(f_L e)$ ($e \in \cG$), the indexes $\sigma_u$ and $\sigma_d$ are used to characterize what generators $g_i \in \cG$ are contained in $e$, and the indexes $i_X, i_Z, j_X, j_Z$ are used for 
recording the changes of the Pauli operators when each such $g_i$ is applied.
Following the graphical representation of the tensor as shown in Fig.~\ref{fig:tensornet}(a), in the following, we may refer to $(i_X, i_Z)$, $(j_X, j_Z)$, $\sigma_u$, and $\sigma_d$ at left, right, top, and bottom indexes of the tensor, respectively.

A tensor $\tilde{p}^{(j)}$, which is dependent on $f_L$ (although not explicitly in our notation), is associated with the Pauli error probabilities $p_I, p_X, p_Y, p_Z$ occurring at the $j$th qubit. This tensor has two left-pointing binary indexes $(i_X, i_Z)$ and has entries from $\{p_I, p_X, p_Y, p_Z\}$. In the case of $f_L=I$, the entries of the tensors $\tilde{p}^{(j)}_{(i_X, i_Z)}$ equal $p_I, p_X, p_Y, p_Z$ for $(i_X, i_Z)=(0,0), (1,0), (1,1), (0,1)$, respectively. For non-trivial $f_L$, one simply flips input bits of $\tilde{p}^{(j)}$ according to the action of $f_L$ on site $j$, so if $f_L^{(j)}=X$, one flips only the $i_X$ bit in the definition above, and if $f_L^{(j)}=Z$, one flips $i_Z$ and for $f_L^{(j)}=Y$ one flips both bits.

We now explain how to determine the layout of the 2D network. The number of rows in the network is always $n_{\rm phys}$, but the number of columns depends on the details of the checks, which we will explain later.
Given a single check, we construct part of the network corresponding to that check as follows. When the check acts as a non-identity Pauli operator $P$ on the $j$th qubit, we place $T^P$ tensor in the $j$th row. We do this for each qubit that the check acts non-trivially on and connect the $\sigma_{u/d}$ indices of all the $T^P$ tensors by a vertical wire. The $\sigma_u$ of the top tensor and $\sigma_d$ of the bottom tensor are each connected to a single index $\delta$ tensor, which has two entries both equal to one and is indicated by a small black circle in Fig. \ref{fig:tensornet}(b). We call this connected set of tensors a check subnetwork.

The set of all check subnetworks, where each subnetwork corresponds to a given check, is then sorted into columns so that no pair of subnetworks in a given column overlap on a row. The right index of every $T^P$ tensor is connected to the left index of the tensor on its right.
The left index of the left-most tensor $T^P$ in each row is fixed to $0$. The right index of the right-most tensor $T^P$ in the $j$th row is connected to the index of $\tilde{p}^{(j)}$. See Fig.~\ref{fig:tensornet}(b) for a specific example of the TN.

The number of columns depends on the choice of checks and can be $n - k$ when there is a qubit shared by all the checks. However, in the case of  1D encoding circuits of depth $d$, the number of columns can be $O(d)$ as the checks can only act non-trivially on at most $2d$ neighboring qubits.

In the 2D TN, each horizontal row is associated with a single physical qubit, and every horizontal wire actually contains two single-bit wires, which we can represent as a single tensor edge of dimension 4. In the $j$th row, the $i_X$ bit corresponds to an $X$ error on the $j$th qubit, and the other $i_Z$ corresponds to a $Z$ error on the $j$th qubit. By contracting with the probability tensor, a probability factor of $p_I^{(j)}, p_X^{(j)}, p_Y^{(j)}$ or $p_Z^{(j)}$, which does not depend on $j$ if the noise is identical for all qubits, will appear, depending on bit values of the wire. Note that the probability tensor is permuted according to $f_L$, as we have explained.

In contracting the TN, it is important to notice that the set of $T^P$ tensors in the network constitutes a large logic circuit and that only valid computations of the classical logic circuit will be summed over (invalid computations evaluate to zero). In valid computations of the classical circuit, the bits $\sigma_{u/d}$ along all vertical wires in the tensors associated with single check qubits must all be 0 or all be 1. 
If $\sigma_{u/d}^{(i)}=1$ for a given check $i$, then the bits on the horizontal wires matching that check will be flipped; otherwise, they will not be affected.
Fixing the vertical check indices $\sigma^{(1)}_{u/d}, \sigma^{(2)}_{u/d}, \dots, \sigma^{(n)}_{u/d}$ and summing over the remaining indices, we can pull out a product of Pauli probabilities from the $\tilde{p}$ tensors, which is equal to $p(f_Le(\sigma))$, with $\sigma=\sigma^{(1)}_{u/d}, \sigma^{(2)}_{u/d}, \dots, \sigma^{(n)}_{u/d}$. Finally, in contracting the TN by summing over the check indices, we sum over all possible configurations $\sigma$ of the check bits and obtain the coset probability $p(f_L \mathcal{G})=\sum_\sigma p(f_Le(\sigma))$.

\subsection{Contracting the network}
The TN can be contracted in various ways. The task is, essentially, to contract an $L\times W$ sized square-lattice TN, where $L$ and $W$ are the length and width, respectively, of the network. To perform the contraction exactly, we first contract the first row into a single tensor with $\mathcal{O}(W)$ indices. We then contract the remaining tensors one by one with this tensor, starting with the first tensor in the second row, then moving down the network along rows then down columns until all tensors are contracted. Assuming $L
\ge W$, the maximum memory cost of this procedure is $\mathcal{O}(2^W)$, and the time cost is $\mathcal{O}(WL2^W)$. For the low-depth random circuits we consider, $L=n_{\rm phys}$ and $W=\mathcal{O}(d)=\mathcal{O}(\log{n_{\rm phys}})$, and so both the time cost and memory cost are polynomial in the block size. 

One could try to improve this scaling further by, e.g., using approximate contraction strategies. The approximate boundary matrix product state (MPS) method, described in Ref.~\cite{schollwock_density-matrix_2011} and used in other decoders~\cite{bravyi_efficient_2014, darmawan_linear-time_2018,chubb_statistical_2021, chubb_general_2021}, would reduce the time and memory costs to  polynomial in $W$ if a truncated bond dimension $\chi$ is kept constant in $N$. We have tested this method, but it appears that the quality of the approximation varies considerably among the codes sampled. A direction of future research could be to find methods to contract the network more efficiently; however, for this work, we obtain our results using the exact contraction method, which is sufficiently fast for our purposes.

\subsection{Decoding logical qubits independently by marginalization}
\label{s:marginal}
The coset probability $p(f_L\mathcal{G})$ can be computed for any $L$ using this method. One problem when encoding a large number of qubits $k$ is that the number of inequivalent logical operators and cosets grows as $4^k$. Therefore finding the most likely coset by computing the probability of every coset takes exponential time in the number of encoded qubits. We overcome this issue by decoding each logical qubit independently. 

To decode a single logical qubit $j$, we calculate the probabilities of the cosets for $f_L \in \{ f, f U l^x_jU^{\dagger}, f Ul^z_j U^{\dagger}, f Ul^x_jl^z_j U^{\dagger}\}$ and marginalize over the other logical qubits, where $l^p_j$ are the unencoded logical Pauli operators for the $j$th logical qubit, and $U$ is the encoding circuit. This marginalization is achieved simply by adding the logical generators $U l^x_i U^{\dagger}$, $U l^z_i U^{\dagger}$ for all $i$ different from $j$ to the list of checks when we construct the TN. In the case of the one-dimensional 1D encoding circuit with depth $d$, adding the logical operators to the TN adds an additional $\mathcal{O}(rd)$ columns to the network and therefore does not change the scaling of the width of the network with $d$. However, the cost of marginalization on exact TN contraction does increase with $r$, which explains why our simulations can only deal with $d=7$ with $r\ge1/3$, while $d=8$ for $r\le1/5$. 

This approach returns the optimal correction for any given logical qubit, but may not be globally optimal due to possible correlations between errors on logical qubits. However, as described in Appendix \ref{s:short_range}, we have observed correlations between logical qubit errors to be short-range.
Based on this observation, we do not expect a significant loss in performance by ignoring correlations between logical qubit failures. 

A TN must be contracted for each coset of each logical qubit, so $4k$ contractions must be evaluated in total. Fortunately, most of the contraction (i.e., the contraction of tensors on qubits on which the logical does not act) can be reused, and thus decoding every qubit requires only a small additional cost compared to decoding a single logical qubit.

\section{Short-range correlations}
\label{s:short_range}
\begin{figure}[t]
    \centering
    \includegraphics{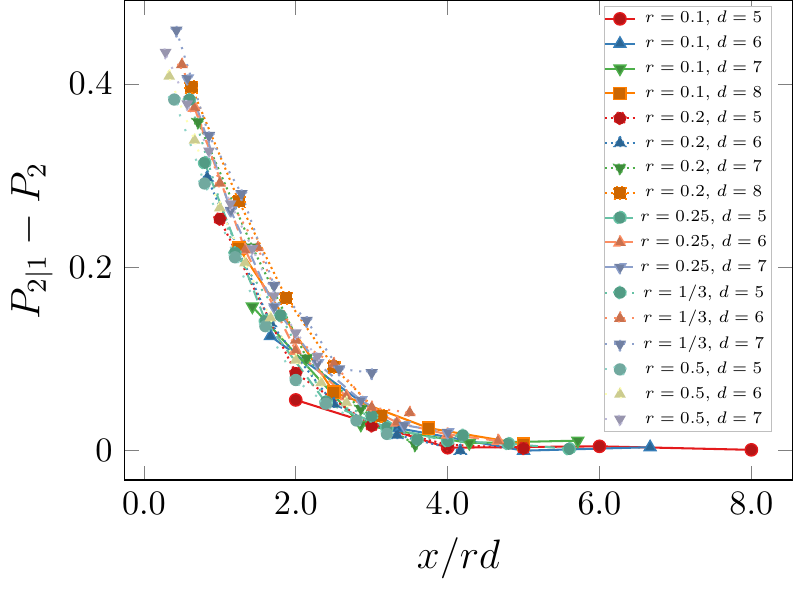}
    \caption{Two-body correlations between qubit failures vs.\ distance normalized by $rd$. The curves appear to collapse onto a single line.}
    \label{fig:correlations}
\end{figure}

We observe that correlations between failures for log-depth circuits under Pauli noise are short-ranged, as was observed in Ref.~\cite{gullans_quantum_2021} for erasure noise. Let $P_{2|1}$ be the probability that qubit 2 fails  given that qubit 1 fails, and $P_{2}$ the probability that qubit 2 fails.
In   Fig.~\ref{fig:correlations}, we have plotted the difference of these quantities, i.e., to what extent the failures are correlated, as a function of the separation $x$ between qubit 1 and 2, normalized by $R$ and $d$ and averaged over all qubit-2 locations. It can be seen that, with this normalization, the various curves collapse, and correlations have a finite range on the order of $rd$. This indicates that, unlike fully random codes, these codes retain some aspects of the spatial locality under local Pauli noise. 

\section{Supplementary numerical results}
\label{s:supplementary_numerics}
In this section, we provide additional results and details on the numerical methods. The thresholds in this paper are estimated using the critical exponent method similar to that described for the surface code in Ref.~\cite{wang_confinement-higgs_2003}, which assumes that, near the threshold, the logical failure rate depends only on the rescaled variable $x=(p-p_c)d^{-1/\nu}$, where $p_c$ is the threshold, and $\nu$ is some critical exponent. Note for the codes with low-depth encoding circuits, we have replaced the lattice dimension (used for the surface code) with the circuit depth $d$. By fitting the bulk logical error probability $p_L'$ to a quadratic polynomial in $x$
\begin{equation}
    p_L'=A+Bx+Cx^2\,,
    \label{e:fit}
\end{equation}
we obtain $p_c$, $\nu$, $A$, $B$ and $C$ as fit estimates. The obtained threshold values for different $r$ along with the hashing bound are displayed in Table.~\ref{t:fit}. 
\begin{table}
\begin{tabular}{|c|c|c|}
\hline
$r$& $p_c$ (TN) &$p_c$ (hashing)\\
\hline
\hline
1/10 & 0.164(2) &0.16305\\
1/5  & 0.144(3) &0.13854\\
1/4  & 0.125(4) &0.12690\\
1/3  & 0.102(3) &0.10835\\
1/2  & 0.061(3) &0.07439\\
\hline
\end{tabular}
\caption{The column $p_c$ (TN) contains the maximum depolarizing error probabilities for error suppression (threshold) we obtain from simulations with codes defined by log-depth 1D random Clifford circuits using the tensor network decoder and fitting our numerical data to Eq.~\eqref{e:fit}. The parenthesized value indicates the error in the last digit, which is  the standard error obtained for the fit estimate. The column $p_c$ (hashing) contains the probabilities given by the hashing bound for this noise model.}
\label{t:fit}
\end{table}

\begin{figure*}
    \centering
    \includegraphics[width=\textwidth]{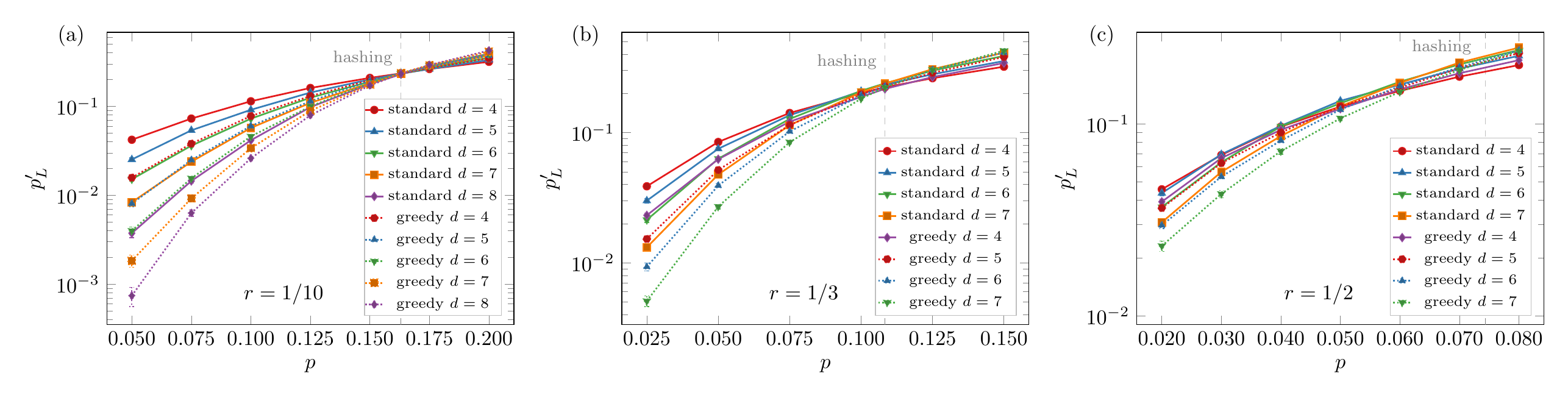}
    \caption{Bulk logical error probability $p_L'$ vs.\ physical error probability $p$ for various encoding circuit depths $d$ for $r=1/10$, $1/3$ and $1/2$ in (a), (b) and (c) respectively. The crossing point of the different curves on each plot corresponds to the threshold. For $r\le 1/3$ the threshold is close to that given by the hashing bound, while a clear crossing point is not discernible for $r=1/2$ for the values of $d$ simulated. The system sizes used for (a), (b), and (c) are $n=50$, $n=54$, and $n=50$ respectively.}
    \label{fig:supplementary_thresholds}
\end{figure*}

We also include additional threshold plots for various rates in Fig.~\ref{fig:supplementary_thresholds}. As can be seen, a clear crossing point is discernible up to $r=1/3$. 

\begin{figure}
    \centering
    \includegraphics[width=0.45\textwidth]{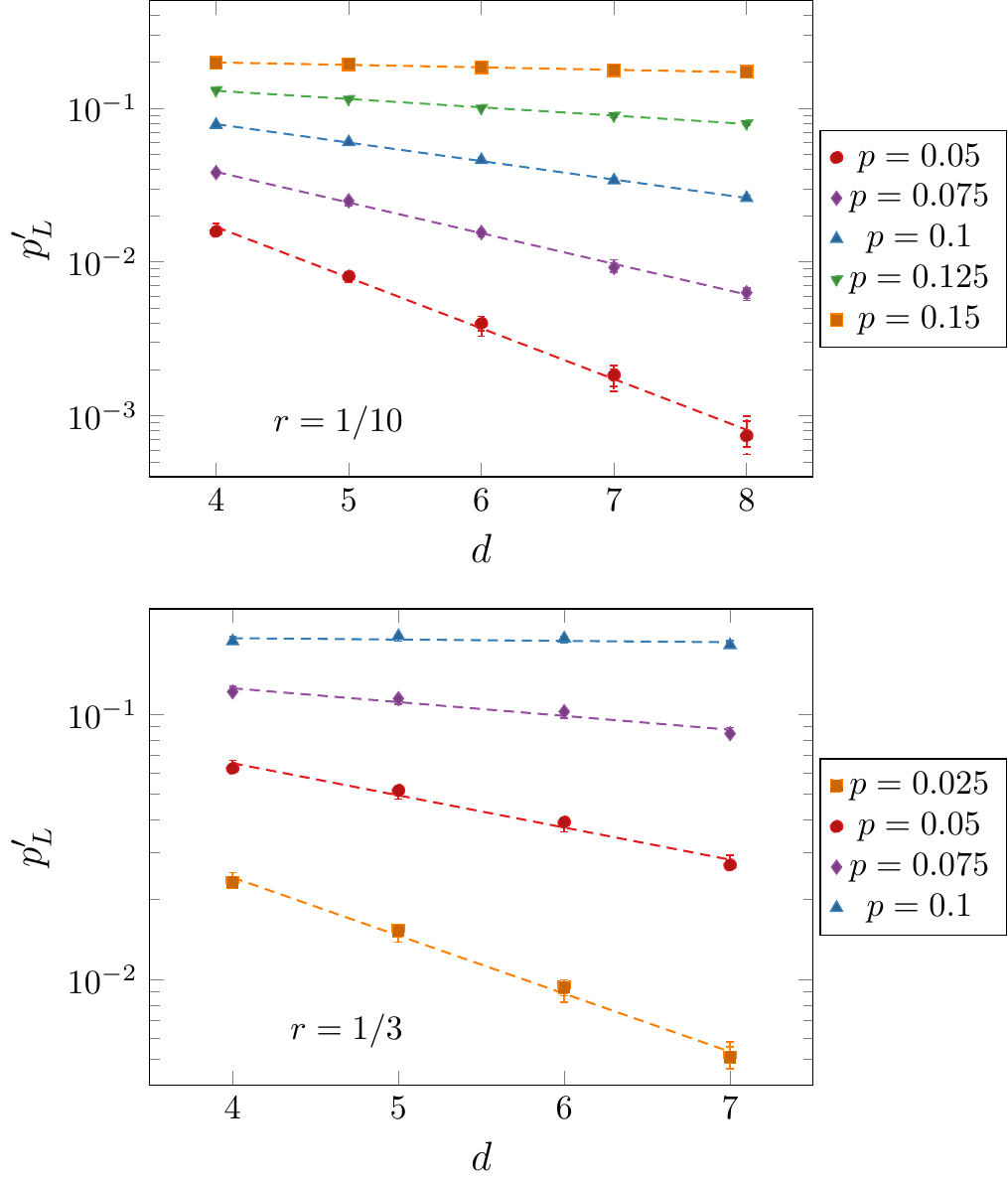}
    \caption{Bulk logical error probability $p_L'$ vs.\ depth $d$ for $r=1/10$ and $r=1/3$ in (a) and (b) respectively, for various error probabilities using the greedy code generator. The dashed lines are obtained by linear regression. A straight-line relationship indicates exponential decay in $d$.}
    \label{fig:exp_decay}
\end{figure}

Finally, in order to illustrate the exponential decay of $p_L'$ as a function of $d$, we have plotted $p_L'$ vs. $d$ on a semi-log plot in Fig. \ref{fig:exp_decay}. A straight-line relationship is observed below the threshold, indicating exponential decay.  

\section{Improvements to fully random Clifford encoding}
\label{s:improvements_random}

Here we describe the `greedy' method to improve the codes produced by random Clifford circuits. This slightly modifies the encoding circuit to avoid low-weight checks and logicals.
The specific details of the Clifford circuit (e.g., the choice of the ${\rm iSWAP}$ as the two-qubit gate) are not expected to make much difference for a high depth, such as $d=\mathcal{O}(n)$, where the generated code is expected to be close to a fully random stabilizer code, as exactly shown for a class of Clifford gate set~\cite{haferkamp_efficient_2022}. However, for a low depth, the choice of gate set can make a large difference, as we will observe in the performance improvement in the `greedy' method.

The first simple modification is to ensure that the initial checks are either single-qubit $X$ or $Y$ before applying the first layer of $\rm iSWAP$ gates. This ensures that the weight of all checks is increased to two by the $\rm iSWAP$ gates (the weight of a single qubit $Z$ would not be changed). After this step, the distance between the first and last sites on which the check acts non-trivially is guaranteed to be increased maximally by two for every layer of $\rm iSWAP$.

Note that, when a two-qubit gate is applied to a Pauli operator acting on 2 qubits, the weight of the operator may increase, decrease or stay the same. Therefore, rather than choosing single-qubit Clifford gates uniformly at random, as in the standard construction, we choose the single-qubit Clifford gate that maximizes the increase in total check weight (sum of the weights of all the checks) when passed through the next $\rm iSWAP$ gate. When multiple gates produce the same increase in weight, one among these is chosen uniformly at random. 

It is harder to avoid low-weight logical operators, since any product of logical generators and stabilizers is also a logical operator, and therefore it is harder to check that the overall weight of logical operators increases or decreases with the application of a two-qubit gate. However, as a simple heuristic, we also add the logical generators to the list of checks whose weight is maximized by greedily choosing single-qubit gates. This appears to further slightly improve performance.

\end{document}